\documentclass[12pt]{amsart}
\usepackage{amssymb}
\usepackage{amscd}
\usepackage{graphics}
\usepackage{verbatim}
\usepackage[usenames]{color}
\usepackage{hyperref}
\usepackage[all]{xy}

\newtheorem{thm}{Theorem}%[section]
\newtheorem{prop}[thm]{Proposition}

\theoremstyle{definition}

\newtheorem{ax}{Axiom}

\theoremstyle{remark}

\newtheorem{rmk}{Remark}

\newenvironment{lsnum}{\begin{enumerate}}{\end{enumerate}}
\newenvironment{pf}{\begin{proof}}{\end{proof}}

\newcommand{\scr}[1]{\ensuremath{\mathcal {#1}}}

\newcommand{\bbb}[1]{\ensuremath{\mathbb {#1}}}

\renewcommand{\phi}{\varphi}

\newcommand{\notarrow}{\kern .42em\not\kern -.42em\longrightarrow}

\newcommand{\ket}[1]{\ensuremath{|#1\rangle}}

\newcommand{\one}[2]{(#1\underset{1}{\cdot}#2)}
\newcommand{\two}[2]{(#1\underset{\tau}{\cdot}#2)}
\DeclareMathOperator{\Hom}{Hom}

\newcommand{\noprint}[1]{\relax}

\title{On Quantum Computation, Anyons, and Categories}
\author{Andreas Blass}
\address{Mathematics Department\\
University of Michigan\\
Ann Arbor, MI 48109--1043, U.S.A.}
\email{ablass@umich.edu}
\author{Yuri Gurevich}
\address{Microsoft Research\\
One Microsoft Way\\
Redmond, WA 98052, U.S.A.}
\email{gurevich@microsoft.com}

\begin{document}

\begin{abstract}
We explain the use of category theory in describing certain sorts of
anyons.  Yoneda's lemma leads to a simplification of that
description.  For the particular case of Fibonacci anyons, we also
exhibit some calculations that seem to be known to the experts
but not explicit in the literature.
\end{abstract}

\maketitle

\section{Introduction}          \label{intro}

This paper attempts to explain the use of category theory in
describing certain sorts of \emph{anyons}. These are rather mysterious
physical phenomena which, one hopes, will provide a basis for quantum
computing needing far less error correction than other approaches.

The first author of this paper has long been a fan of category theory;
even as a graduate student, he was described by one of his professors
as ``functorized''.  The second author has been far more skeptical
about the value of category theory in computer science, because of its
distance from applications and because of the peril of potential (and
in some cases actual) over-abstraction.  In 2012, both authors began
working with the Quantum Architectures and Computing (QuArC) Group at
Microsoft Research and found anyons to be near the top of the group's
agenda.
Seeing calculations and applications that use unitary matrices to
represent braiding of anyons, we naturally wondered what Hilbert space
these matrices are intended to operate on.  We made rather a nuisance of
ourselves by asking different people, on different occasions, what
anyons actually are, from a mathematical point of view. Are they
Hilbert spaces? Are they vectors in a Hilbert space?  Are they
something else?  It turned out that the only mathematically sound
answer in the literature involved a special sort of categories,
\emph{modular tensor categories}.\footnote{Other answers explained the
  physics, in terms of excitations, but these matters are not the
  subject of this paper, which is specifically about mathematics
  except for the introductory material summarized in
  Section~\ref{anyons}.}  So the second author agreed that categories
can be quite relevant to important applications in computer science.

Our purpose in this paper is to describe some of the ideas surrounding
categories and anyons in general and the special case of Fibonacci
anyons and their category description.  We hope that our presentation
will be accessible and useful for mathematicians and computer
scientists who have some acquaintance with the basics of category
theory.  Where we need to go beyond the basics, we explain, albeit
briefly, the concepts from category theory that we use.  We have also
included a section describing the physical background that this
mathematics is intended to formalize.

To describe more of our motivation for studying anyons, we need to
presuppose some general information that will be explained in later
sections of this paper.  In particular, we shall refer to the fusion
rule $\tau\otimes\tau=\tau\oplus 1$ for Fibonacci anyons $\tau$ (and
the vacuum $1$).  We hope that the following paragraphs will give the
reader a rough idea of what we are looking at, and that re-reading
them after the rest of the paper will provide a less rough idea.

In contrast to what occurs elsewhere in quantum theory, the states (represented, as usual, by vectors in Hilbert spaces, up to scalar multiples) in the modular tensor category picture are ways in which one configuration can fuse to form another configuration.\footnote{For more on the notion of fusion, see Remark~\ref{fuse} at the end of this introduction.}  They are not the configurations themselves.  For example, in the Fibonacci case, there is a 2-dimensional Hilbert space of ways for three anyons to be regarded as (or to fuse into) one anyon; this is the Hilbert space $\text{Hom}(\tau\otimes\tau\otimes\tau,\tau)$.

When we first heard about Fibonacci anyons, we thought that the fusion
rule $\tau\otimes \tau = \tau \oplus 1$ meant that, if we put two $\tau$
anyons together, then the result might look like one $\tau$ anyon or
like the vacuum (this much is true in the modular tensor category
model) and that the general result would be a superposition of these
two alternatives.  But the model doesn't allow such
superpositions. Nor does the model say anything about the
probabilities of the two possible outcomes.

Instead, we get superpositions of the following sort.  Start with
three $\tau$'s.  Fuse the first two to get one $\tau$ or vacuum.  If
you got vacuum, then the overall result is one $\tau$, namely the third
of the original ones, which you haven't yet fused.  If, on the other
hand, fusing the first two $\tau$'s gives a $\tau$, then fusing that
with the third $\tau$ might produce a $\tau$.  (It might also produce
vacuum, but that's irrelevant for the present discussion.)  So we have
two ways to end up with one $\tau$, according to whether the first two
$\tau$'s fused to vacuum or to $\tau$.  And it is these two ways that
the model allows superpositions of.  Another possibility for
getting two ways here is to fuse the last two $\tau$'s first and
then fuse the result with the first $\tau$.  These two form another basis
of the same 2-dimensional Hilbert space of ``ways".  The relation
between the two ways is (part of) the associativity isomorphism of the
modular tensor category.  Yet another possibility would begin by
fusing the first and third $\tau$'s.  The modular tensor category
representation of this possibility would use a braiding
isomorphism to move the first anyon to be adjacent to the third (or
vice versa), and it would depend on the path along which that anyon is
moved around the second one.

In Section~\ref{anyons}, we give a general introduction to
anyons from the point of view of physics and quantum computation.
That section is intended to give the reader a rough idea of what
anyons are and why researchers in quantum computation would be
interested in them.  The treatment here is quite superficial, and we
give references for more detailed treatments.

In Section~\ref{mtc}, we gradually introduce modular tensor
categories, and we explain how they are intended to be used to
describe anyons.  This section borrows heavily from the axiomatization
given in \cite{panang}, but with some modifications and
rearrangements.

Section~\ref{yoneda} is devoted to an application of one of the
central theorems of category theory, known as Yoneda's Lemma, to
producing a simplified view of modular tensor categories.

Finally, in Section~\ref{fib}, we consider the special case of
Fibonacci anyons.  This special case is unusually simple in some
respects.  Nevertheless (or perhaps therefore) it occupies a prominent
place in quantum computing research. Section~\ref{fib} begins with a
general description of Fibonacci anyons and then exhibits some
calculations, whose results seem to be well known to some in the
quantum computing community but which we have not been able to find
written down in the literature.

More detailed treatments of modular tensor categories are available in
the papers \cite{panang} of Panangaden and Paquette and \cite{wang} of
Wang.  Much of our exposition is based on the former. For other
aspects of anyons and topological quantum computation, see, for
example, \cite{kl} and the references there.

\begin{rmk}             \label{fuse}
  We encountered numerous explanations of the notion of \emph{fusion}
  of anyons, and they seemed to contradict each other.  At one extreme
  was the picture of fusion as a physical process in which anyons are
  brought into spatial proximity with each other and energy is
  released as they form a new anyon (or perhaps annihilate each
  other).  A minor modification of this picture is that energy need
  not be released; it might actually be consumed in the process.
  Another picture, however, did not insist that the anyons be brought
  together.  They could remain far apart, and a suitable global
  measurement of the system's quantum numbers could reveal how they
  ``fused''.  A path to reconciling these apparently contradictory
  pictures is suggested by a comment at the end of Section~II.A of
  \cite{nssfd}; the idea is as follows. Consider several anyons, which
  we intend to fuse.  As long as they are far apart, the various
  possible results of their fusion have energies that are very close
  together.  (In technical terms, the ground state of the system is
  very nearly degenerate.) So the different fusion results can be
  distinguished in principle but not practically.  When the anyons are
  brought closer together, though, the energy differences between the
  fusion possibilities become larger, and so it becomes practical to
  distinguish these possibilities.  Thus, the discrepancy between
  various views of fusion seems to be largely a discrepancy between
  what can be observed in principle (or what is ``really'' happening)
  and what can be detected in practice.
\end{rmk}

\section{Quantum theory and anyons}                \label{anyons}

This section is a superficial summary of a small part of quantum
theory and some basic information about anyons.  The physics described
here is intended merely to provide an orientation for understanding
the mathematics in the rest of the paper.

\subsection{Quantum Mechanics}

In quantum theory, the state of a physical system is typically
represented by a non-zero vector in a complex Hilbert space \scr H,
but all non-zero scalar multiples of a vector represent the same
state.  Thus, the states constitute the projective space associated to
\scr H.  Because of the freedom to adjust scalar factors, one often
imposes the normalization that the vectors representing a state should
have norm 1; there still remains a freedom to adjust the phase, i.e.,
a scalar factor of absolute value 1.

If a system has an observable property with infinitely many possible
values, for example position or momentum, then the Hilbert space of
its states must be infinite-dimensional.  In quantum computing,
however, one usually ignores many such properties and concentrates on
only a small number (often only one) of properties with only finitely
many possible values. As a result, one deals with finite-dimensional
Hilbert spaces. (This simplification is analogous to modeling a
classical computer by a configuration of bits, not taking account of
its other physical properties, like position or momentum or
temperature, unless these threaten to interfere with the bits of
interest.)

The automorphisms of a Hilbert space \scr H are the \emph{unitary}
transformations, i.e., the linear bijections that preserve the inner
product structure.  These play several important roles, both in
physics and in quantum computation.  First, they provide the dynamics
of isolated quantum systems.  That is, the state of an isolated system
will evolve in time by the action of a one-parameter group (the
parameter being time) of unitary operators.\footnote{Here we use the
  so-called Schr\"odinger picture of quantum mechanics. A physically
  equivalent alternative view, the Heisenberg picture, has the states
  remaining constant in time, while the operators modeling properties
  of the state evolve by conjugation with a one-parameter group of
  unitary operators.}  Second, if a system has symmetries, i.e., if it
is invariant under some transformations, then these transformations
are usually modeled by unitary operators.\footnote{A few discrete
  symmetries can be modeled by anti-unitary transformations.}
Finally, the design of quantum algorithms is based on unitary
operators.  We want the system to evolve from a state that we know how
to produce to another state from which we can extract useful
information by a measurement. That evolution is described by a unitary
operator. So an algorithm designer wants to find unitary operators
that represent a useful evolution of a state.  In addition to finding
such operators, we want to represent them as compositions of simpler
ones, called \emph{gates}, that we know how to implement.

Where classical computation uses bits, whose possible values are
denoted by 0 and 1, quantum computation uses \emph{qubits}.  A
measurement of a qubit produces two possible values; the qubit itself
is represented by a 2-dimensional Hilbert space, in which a certain
orthonormal basis, usually written $\{\ket0,\ket1\}$, corresponds to
the two values.  In contrast to the classical case, though, the
Hilbert space structure provides many other states in addition to
these two basic ones.  Any non-zero linear combination of \ket0 and
\ket1 represents a possible state of the system.  If the state is
represented by the unit vector $x\ket0+y\ket1$, then measuring the
qubit in the $\{\ket0,\ket1\}$ basis will produce the outcome 0 with
probability $|x|^2$ and the outcome 1 with probability $|y|^2$.  Such
a state is a \emph{superposition} of the two basic states.  More
precisely, this state vector is the superposition, with coefficients
$x$ and $y$, of the vectors $\ket0$ and $\ket1$, respectively.

It is more accurate to speak of superposition of vectors than of
superposition of states.  The reason is that, although phase factors
don't affect the state represented by a vector, \emph{relative} phases
do affect superpositions.  Thus, for example, although $\ket1$ and
$-\ket1$ represent the same state of a qubit, the superpositions
$(\ket0+\ket1)/\sqrt2$ and $(\ket0-\ket1)/\sqrt2$ represent quite
different states.

It is almost true in general that, for any two states of any quantum
system, any superposition of the associated vectors also represents a
possible state of that system.  The word ``almost'' in the preceding
sentence refers to the possibility of \emph{superselection
  rules}. These rules specify that, for certain quantities, like
electric charge, it is impossible to superpose two states with
different values of those quantities.  Thus, when discussing a system
for which several values of the electric charge can occur, we are, in
effect, dealing with several separate Hilbert spaces, called
\emph{superselection sectors}, one for each value of the charge.  One
can, and sometimes one does, form the direct sum of these Hilbert
spaces to obtain a Hilbert space containing all the possible states of
that system, but most of the vectors in that direct sum, involving
superpositions with different charges, do not represent physically
possible states.  We prefer, in this paper, to deal with
superselection sectors as separate Hilbert spaces and forgo their
direct sum.  For more information about superselection rules, see
\cite{giulini}.

In reality, there are very few superselection rules --- arising from
certain conserved quantities like electric charge, baryon number, and
parity --- but in the study and application of anyons one often
artificially adds superselection rules, and we shall encounter such
rules in the category-theoretic treatment below.  This amounts to
deciding not to consider superpositions of vectors from certain
Hilbert spaces, i.e., to consider those superselection sectors
separately rather than considering their direct sum.

In the presence of superselection rules, the operators that one
considers are operators acting on each of the superselection sectors
separately. In the case of true superselection rules, the dynamics of
the system and any gates that one could construct are given by
unitary operators acting on each sector separately. In the case of
artificial superselection rules, nature may not cooperate with our
artificial rules, and states in one sector may evolve out of that
sector.  Such evolution interferes with our understanding
and intentions; it is often called ``leakage'' and one strives to
avoid it.

In addition to the unitary operators mentioned above, Hermitian (or
self-adjoint) operators on the Hilbert space of states also play an
important role in quantum mechanics, because they model observable
properties of a system.  The connection between Hermitian operators
and (real-valued) observables is easy to describe in the case of
finite-dimensional Hilbert spaces \scr H.\footnote{In the
  infinite-dimensional case, the description is similar but one must
  take into account the possibility of a continuous spectrum of the
  operator, in addition to or instead of discrete eigenvalues.}  Let
the Hermitian operator $A$ have (distinct) eigenvalues
$a_1,\dots,a_k$, with associated eigenspaces $S_1,\dots,S_k$. (Some of
these eigenvalues may have multiplicity greater than 1, but they are
to be listed only once among the $a_i$'s. The associated $S_i$ will
then have dimension greater than 1.)  These eigenspaces are orthogonal
to each other, and their sum is all of \scr H. Any unit vector
$\ket\psi\in\scr H$ can be expressed as the sum of its projections
$\ket{\phi_i}$ to the subspaces $S_i$.  Measuring $A$ on a system in
state $\ket\psi$ produces one of the eigenvalues $a_i$; the
probability of getting the result $a_i$ is the squared norm of the
projection, $\Vert\ket{\phi_i}\Vert^2$.  Note that the dimension of
\scr H is an upper bound for the number of distinct eigenvalues $a_i$
of any Hermitian operator on \scr H.  In particular, any measurement
performed on a qubit will have at most two possible outcomes.  It is
in this sense that a qubit is the quantum analog of a classical bit.

\subsection{Anyons}     \label{subsec:anyons}

To understand anyons, it is useful to recall first that ordinary
particles are of two sorts, \emph{bosons} and \emph{fermions}. These
differ in several respects, beginning with the action of spatial
rotations on the corresponding Hilbert spaces. For particles in
ordinary 3-dimensional space, the group $SO(3)$ of Euclidean rotations
of that space acts on the states of the particle.  (More precisely,
the group of all Euclidean motions acts, but we abstract from the
particle's position and consider only its orientation in space; thus
we ignore translations and consider only the group of rotations.)
Because the vector representing a state is defined only up to a phase
factor, the action of the rotation group is not a representation in
the usual sense but a \emph{projective representation}.  This means
that each rotation $g$ of physical 3-dimensional space is represented
by a unitary operator $\rho(g)$ on the Hilbert space, but this
$\rho(g)$ is unique only up to a phase factor.  It is customary to
make some arbitrary choice of these phase factors, so that we can
speak unambiguously of $\rho(g)$.  The arbitrariness of the choice is,
however, reflected in the fact that $\rho(gh)$ and
$\rho(g)\rho(h)$ need not be equal but can differ by a phase factor.
Furthermore, $\rho$ and $\rho'$ are considered equivalent
representations if they differ only by these arbitrary phase factors.
It is reasonable to ask, in this connection, why the operators
$\rho(g)$ need to be unitary or even linear, rather than only linear
up to phase factors.  The reason is that, unlike absolute phases,
relative phases are relevant in superpositions, so physical symmetries
must preserve them.

It turns out that any projective representation $\rho$ of $SO(3)$ is
given by a genuine unitary representation $\tilde\rho$ of the
universal covering group of $SO(3)$, namely $SU(2)$ (see for example
\cite{b-m} and \cite{Raghu}).  That is, if $p:SU(2)\to SO(3)$ denotes
the 2-to-1 projection map, we have $\rho\circ p$ equivalent to
$\tilde\rho$.  More concretely, it means that there are two sorts of
projective representations of $SO(3)$, up to equivalence.  One sort is
the ordinary unitary representations of $SO(3)$; the other is given by
unitary representations of $SU(2)$ that send the non-trivial element
$-I$ of the kernel of $p$ to the operator $-I$.  (Throughout this
paper, we use $I$, sometimes with subscripts, to denote identity
transformations, functions, morphisms, etc.)  The first sort of
representation corresponds to bosons, whose state vectors (not merely
their states) are unchanged when rotated gradually through a full
revolution.  The second sort corresponds to fermions, where a rotation
through $2\pi$ changes the state vector by a sign.

A second distinction between bosons and fermions, even more important
for our purposes, is the behavior of systems of several identical
particles.  Because the particles are identical, any permutation of
the particles leaves the state unchanged and therefore changes the
state vector by at most a phase factor.  As a result, we have a
one-dimensional projective representation of the symmetric group.
Again, it turns out that there are just two possibilities (both of
which are actual unitary representations of the symmetric group).
Either all permutations leave the state vectors unchanged, or the even
permutations leave the state vectors unchanged while the odd
permutations reverse the vectors' signs.

A deep theorem of relativistic quantum field theory, the
spin-statistics theorem, says that these two behaviors of
multi-particle states under permutations exactly match the two
behaviors of single-particle states under rotations.  Interchanging
two identical bosons leaves the state vector of the pair unchanged;
interchanging two identical fermions reverses the sign of the state
vector.

The preceding discussion of bosons and fermions depends crucially on
the fact that the particles are in ordinary 3-dimensional space.  If
particles were confined to a 2-dimensional space, more possibilities
would arise.

Specifically, the rotation group in two dimensions, $SO(2)$, has more
sorts of projective representations than $SO(3)$ does; the reason is
ultimately that the universal covering group of the circle group
$SO(2)$ is the additive group of real numbers, and the covering
projection is not 2-to-1 but $\infty$-to-1.  The result is that a
gradual rotation of a particle through $2\pi$ can multiply its state
vector by an arbitrary phase factor, not just $\pm1$.  The possibility
of getting \emph{any} phase here led to the name \emph{anyon}.

Reducing the dimensionality of space from 3 to 2 also affects the
possibilities for permuting identical particles.  For simplicity,
consider the case where there are just two particles, and we
interchange them.  We can perform the intechange gradually, in the
plane, by rotating the 2-particle system counterclockwise by $\pi$
around the midpoint between the particles.  Alternatively, we can
achieve the same interchange by a clockwise rotation.  In
3-dimensional space, these two options are equivalent in the sense
that they can be gradually deformed into each other, by rotating the
plane of the particles' motion about the line through the particles'
initial positions.  In 2-dimensional space, there is no such
deformation without making the particles collide.  Winding one
particle around the other any number of times, we get infinitely many
ways to achieve one and the same permutation.  With more than two
particles, there are more complicated ways to achieve the same
permutation by moving the particles around in the plane.
As a result, in place of (projective) representations of
symmetric groups, we have representations of braid groups.  For
example, in the case of two particles, in place of the group of two
possible permutations of the particles, we have the group of all
integers, with integer $n$ representing a counterclockwise rotation by
$n\pi$ (and negative $n$ representing clockwise rotations).

The preceding discussion was oversimplified in that (among other
things) when moving particles around each other, we ignored any
rotation that the individual particles might have undergone during the
motion.  A more accurate presentation would need to suitably combine
the braid and rotation groups.

\subsection{Anyons in Reality}

As explained above, anyons do not occur in 3-dimensional space; it is
necessary to reduce the number of spatial dimensions to 2.  Since we
live in a 3-dimensional space, will we ever find anyons?  It turns out
that anyon-like behavior occurs for certain excitations in materials
that are so thin as to be effectively two-dimensional.  A detailed
discussion of this would take us too far from the purpose of this
paper, so we refer the reader to Section~1.1 of \cite{panang}.

We emphasize, however, that the anyons are not what one would
ordinarily think of as ``particles'' but rather excitations in some
medium, which exhibit particle-like behavior.  It should be noted in
this connection that it is not unusual, in other contexts, for
excitations to behave like particles and thus to be analyzed
mathematically as if they were particles.  For example, vibrational
excitations in crystal lattices are treated as particles called
phonons.  Similarly, photons are excitations of the electromagnetic
field.  In quantum field theory, all particles are excitations of the
corresponding fields.

\subsection{Anyons in Quantum Computation}
Quantum computation is unpleasantly susceptible to environmental
disturbances.  Its advantages over classical computation depend on
maintaining superpositions of state vectors, with high precision in
the coefficients of those vectors.  Small disturbances can easily
modify those coefficients or, indeed, destroy superpositions
altogether.  Significant effort must therefore be devoted to error
correction, and this makes algorithms slower and harder to design.

It has been suggested \cite{kitaev} that qubits could be more robust,
i.e., less susceptible to disturbances, if they were implemented using
certain sorts of anyons.
For example, if qubits were encoded in the way two
anyons wind around each other, then this winding, being a topological
property of the system, would be robust.  A small disturbance in the
actual motion of the anyons would leave the winding number intact.
This hope of reducing the error correction needs of quantum computing
has motivated much of the current interest in anyons.

In this approach to quantum computation, braiding of anyons serves not
only to store information but also to process it.  In general, as
mentioned above, quantum computation proceeds by initializing a
quantum state, then applying a unitary transformation to it, and
finally measuring some observable in the resulting transformed state.
The unitary transformation used here must be designed so that a
feasible measurement produces a useful result.  In addition, there
must be a way to implement the unitary transformation as the
composition of a sequence of simpler unitary tranformations, usually
called gates in this context.  In the anyon approach to quantum
computation, the most basic unitary gates arise from the braiding of
anyons around each other, and a crucial question is whether these
gates are \emph{universal} in the sense that arbitrary gates can be
approximated by composing the basic ones.

It is worth noting explicitly that, in this picture, a qubit is not
encoded in the state of a single anyon but rather in a whole system of
several anyons.  This feature will be quite prominent in the category
picture described in the rest of this paper.

\section{Modular tensor categories}     \label{mtc}

In this section we describe the category-theoretic structure that has
been developed to support a mathematical theory of anyons.  Much of
what we describe here is in \cite{panang}, though we have modified
some aspects and rearranged others.

Throughout this section, we let \scr A be a category, intended to
describe the quantum-mechanical behavior of a system of anyons.
\scr A will carry several sorts of additional structure, roughly
classified as ``additive'' and ``multiplicative'' structure, all
subject to various axioms.  We describe the structures and the axioms
a little at a time.  We begin with the additive structure, because
this is where Hilbert spaces enter the picture, so it is the basis for
the connection with the usual formalism of quantum theory.

The vectors in our Hilbert spaces will be the morphisms of \scr A.
Specifically, for each pair of objects $X,Y$ of \scr A, the set
$\Hom(X,Y)$ of morphisms from $X$ to $Y$ will have the structure of a
Hilbert space.  So we have many Hilbert spaces, one for each pair
$X,Y$ of objects.  Some of these Hilbert spaces will be mere
combinations of others, but there will still be several different
``basic'' Hilbert spaces.  This means physically that we regard the
system as being subject to superselection rules, which keep these
Hilbert spaces separate.

We assume familiarity with some basic notions of category theory,
specifically, the notions of product (including terminal object, which
is the product of the empty family), coproduct (including initial
object), equalizer, coequalizer, monomorphism, epimorphism,
isomorphism, functor, and natural transformation. Definitions and
examples can be found in \cite{maclane} or \cite[Chapter~1]{freyd}.

\subsection{Additive Structure}

We begin by requiring \scr A to be an abelian category.  This
requirement, formulated in detail below, provides a well-behaved
addition operation on each of the sets $\Hom(X,Y)$, although the
requirement is formulated in purely category-theoretic terms and does
not explicitly mention this addition operation.

\begin{ax}[\textbf{Abelian}]
\scr A is an abelian category.  That is
\begin{lsnum}
\item There is an object 0 that is both initial and terminal.  A
  morphism that factors through this zero object will be called a zero
  morphism and denoted by 0. Note that each $\Hom(X,Y)$ contains a
  unique zero morphism.
\item Every two objects have a product and a coproduct.
\item For every morphism $\alpha:X\to Y$, the pair $\alpha,0$ has an
  equalizer and a coequalizer.  These are called \looseness=-1 the
  \emph{kernel} and \emph{cokernel} of $\alpha$.
\item Every monomorphism is the kernel of some morphism, and every
  epimorphism is the cokernel of some morphism.
\end{lsnum}
\end{ax}

This axiom has a surprisingly rich collection of consequences,
developed in detail in Chapter~2 of \cite{freyd}.  We list here only
some of the highlights, which will be important for this paper, and we
refer the reader to \cite{freyd} for the proofs and additional
information.

\begin{prop}[\cite{freyd}, Theorem~2.12]        \label{mono-epi}
Any morphism that is both monic and epic is an isomorphism.
\end{prop}

(More generally, as one can easily check, in any category, any equalizer that is an
epimorphism is an isomorphism.)

\begin{prop}[\cite{freyd}, Theorem~2.35]
The product and coproduct of any two objects coincide.
\end{prop}

That is, given two objects $X$ and $Y$, there is an object $X\oplus Y$
that serves simultaneously as the product of $X$ and $Y$, with
projections $p_X:X\oplus Y\to X$ and $p_Y:X\oplus Y\to Y$, and as the
coproduct of $X$ and $Y$, with injections $u_X:X\to X\oplus Y$ and
$u_Y:Y\to X\oplus Y$.  (If $X=Y$, then our notations for the
projections and injections become ambiguous, and we use
$p_1,p_2,u_1,u_2$ instead.) For brevity, we often refer to $X\oplus Y$
as the \emph{sum} of $X$ and $Y$, rather than as the product or
coproduct.

As a product, $X\oplus X$ admits a diagonal morphism $\Delta_X: X\to
X\oplus X$, namely the unique morphism whose composites with both
projections are the identity morphism $I_X$ of $X$.  Dually, as a
coproduct, it admits the folding morphism $\nabla_X:X\oplus X\to X$,
whose composites with both of the injections are $I_X$.  Using the
diagonal and folding morphisms, we can define a binary operation,
called addition, on $\Hom(X,Y)$ for any objects $X$ and $Y$.  Given
$f,g:X\to Y$, we define $f+g:X\to Y$ to be the composite
\[
X\overset{\Delta_X}{\longrightarrow}X\oplus X
\overset{f\oplus g}{\longrightarrow}
Y\oplus Y\overset{\nabla_Y}{\longrightarrow}Y,
\]
where $f\oplus g$ is obtained from the functoriality of products (or
of coproducts --- they yield the same result).

\begin{prop}[\cite{freyd}, Theorems~2.37 and 2.39]
This addition operation makes each $\Hom(X,Y)$ an abelian group, with
the zero morphism serving as the identity of the group.  Composition
of morphisms is additive with respect to both factors; that is, when
either factor is fixed, the composite $f\circ g$ is an additive
function of the other factor.
\end{prop}

\begin{ax}[\textbf{Vectors}]
  Each of these abelian groups $\Hom(X,Y)$ carries an operation of
  multiplication by complex numbers, making $\Hom(X,Y)$ a vector space
  over \bbb C, and making composition of morphisms bilinear over \bbb
  C.
\end{ax}

The complex vector spaces $\Hom(X,Y)$ will play the role of
quantum-mechanical state spaces.  For this purpose, they should also
be equipped with inner products, making them Hilbert spaces, but,
following \cite{panang}, we refrain from assuming an inner product
structure at this stage of the development.\footnote{In fact, inner
  products are never explicitly assumed in \cite{panang}.  They are,
  however, implicit in the statement, in Section 5.1 of \cite{panang},
  that certain bases ``are -- of course -- related by a unitary
  transformation''.}  It turns out that much of what we shall do later
does not depend on the availability of inner products in the vector
spaces $\Hom(X,Y)$.

An object $S$ in the abelian category \scr A is called \emph{simple}
if $S\not\cong0$ and every monomorphism into $S$ is either a zero
morphism or an isomorphism.  In other words, $S$ is a non-zero object
with no non-trivial subobjects.  Because of the abelian structure of
\scr A, this definition can be shown (using
\cite[Theorem~2.11]{freyd}) to be equivalent to its dual: A non-zero object
is simple if and only if it has no non-trivial quotients, i.e., every
epimorphism out of $S$ is either a zero morphism or an isomorphism.

\begin{ax}[\textbf{Semisimple}]
Every object in \scr A is a finite sum of simple objects.
\end{ax}

This axiom considerably simplifies the structure of the vector spaces
$\Hom(X,Y)$.  In the first place, as shown in
\cite[Section~2.3]{freyd}, morphisms from a sum $\bigoplus_jS_j$ to
another sum $\bigoplus_kS'_k$ are given by matrices of morphisms
between the summands.  Specifically, the matrix associated to
$f:\bigoplus_jS_j\to\bigoplus_kS'_k$ has as its $a,b$ entry the
composite
\[
S_b \overset{u_b}\longrightarrow \bigoplus_jS_j
\overset{f}\longrightarrow\bigoplus_kS'_k
\overset{p'_a}\longrightarrow S'_a.
\]
Composition of morphisms in \scr A corresponds to the usual
multiplication of matrices.

Furthermore, when the summands are simple, we have the following
additional information about the matrix entries, a generalization of
Schur's Lemma in group representation theory.

\begin{prop}    \label{schur}
If $f:S\to S'$ is a morphism between two simple objects, then $f$ is
either the zero morphism or an isomorphism.
\end{prop}

\begin{pf}
The kernel of $f$ is a monomorphism into $S$, and if it is an
isomorphism then $f$ is zero. So, by simplicity of $S$, we may assume
that the kernel of $f$ is zero and therefore (by
\cite[Theorem~2.17*]{freyd}) $f$ is a monomorphism.
Similarly, by considering the cokernel of $f$ and invoking the
simplicity of $S'$, we may assume that $f$ is an epimorphism.  But
then, by Proposition~\ref{mono-epi}, $f$ is an isomorphism.
\end{pf}

The last axiom in this subsection combines two finiteness assumptions.

\begin{ax}[\textbf{Finiteness}]
  \begin{lsnum}
    \item There are only finitely many non-isomorphic simple objects.
\item Each of the vector spaces $\Hom(X,Y)$ is finite-dimensional over
  \bbb C.
  \end{lsnum}
\end{ax}

The first of these two finiteness requirements is merely a technical
convenience.  The second, however, gives the following important
information about the endomorphisms of simple objects.

\begin{prop}
  If $S$ is a simple object, then $\Hom(S,S)\cong\bbb C$.
\end{prop}

\begin{pf}
The operation of composition of morphisms is a multiplication
operation that makes the vector space $\Hom(S,S)$ into an algebra over
\bbb C.  Since $S$ is simple, Proposition~\ref{schur} says that every
non-zero element of this algebra is invertible.  That is, $\Hom(S,S)$
is a division algebra over \bbb C.  But \bbb C is algebraically
closed, so the only finite-dimensional division algebra over it is
\bbb C itself.
\end{pf}

Note that the isomorphism $\Hom(S,S)\cong\bbb C$ in this proposition
can be taken, as the proof shows, to be an isomorphism of algebras,
not just of vector spaces.  In particular, the identity morphism of
$S$ corresponds to the number 1.

Combining this proposition with our earlier observations about
matrices, we find that any morphism
$f:\bigoplus_jS_j\to\bigoplus_kS'_k$ between any two objects in \scr
A is given by a matrix whose entries are complex numbers.  Moreover,
the $a,b$ entry is 0 unless $S_b\cong S'_a$.  From this observation,
it easily follows that, when an object $X$ of \scr A  is expressed as
a sum $\bigoplus_jS_j$ of simple objects, the isomorphism types of the
summands $S_j$ and their multiplicities are completely determined by
$X$.  That is, the representation of $X$ as a sum of simple objects is
essentially unique.

\subsection{Multiplicative Structure}
In this subsection, we introduce the multiplicative structure that makes
\scr A a braided monoidal category.  The central idea is that, if
objects $X$ and $Y$ represent certain anyons, then $X\otimes Y$ should
represent a system consisting of both of these anyons.  We must,
however, remember that the Hilbert spaces that occur in this context
are not the objects of \scr A but the vector spaces of morphisms
between the objects.

A system consisting of two anyons of types $X$
and $Y$ would, if measured as a whole, appear as another anyon, whose
type might not be entirely determined by the types $X$ and $Y$.
Formally, this means that $X\otimes Y$ is a sum of several simple
objects.  Furthermore, there might be several ``ways'' for a composite
system to appear as having a particular type $Z$, modeled as several
morphisms from $X\otimes Y$ to $Z$, and our Hilbert spaces will also
contain superpositions of these.

The multiplicative structure will also include a unit object 1; its
intended interpretation is the vacuum.  Thus, $1\otimes X$ and
$X\otimes 1$ amount to just $X$ because a system consisting of $X$ and
nothing is the same as $X$.

The first aspect of multiplicative structure can be stated rather
briefly as the following axiom, but we expand it afterward because we
shall need  the details later.

\begin{ax}[\textbf{Multiplication}]
\scr A is a monoidal category.
\end{ax}

This means that it is equipped with a ``multiplication'' functor
$\otimes:\scr A\times\scr A\to\scr A$ and a ``unit object'' 1 that
satisfy the usual associative and unit laws up to coherent
isomorphism.  Let us first explain ``satisfying the laws up to
isomorphism'' and then discuss ``coherent''.

Associativity would mean that $A\otimes (B\otimes C)$ is the same as
$(A\otimes B)\otimes C$ for any objects $A,B,C$ (and similarly for
morphisms).  Associativity up to isomorphism means that these objects
need not be equal but they are isomorphic, and we are given specific
isomorphisms
\[
\alpha_{A,B,C}:(A\otimes B)\otimes C\to A\otimes (B\otimes C)
\]
for all $A,B,C$, and furthermore  these isomorphisms constitute a
natural transformation between functors $\scr A\times\scr A\times\scr
A\to\scr A$.

Similarly, the requirement that the object 1 be a unit up to
isomorphism means that we are given natural isomorphisms
\[
\lambda_A:1\otimes A\to A\quad\text{and}\quad
\rho_A:A\otimes 1\to A.
\]

As is well-known from classical algebra, the associative law implies
associative identities for more than three factors at a time; for
example, if $*$ is an associative operation, then all five of the
possible parenthesizations of $a*b*c*d$ give the same result.  The
analogous result for categories is that any natural isomorphism
$\alpha$ as above produces natural isomorphisms between any two
parenthesizations of $A\otimes B\otimes C\otimes D$.  There is,
however, an embarrassment of riches, as we can build, from $\alpha$
(and its inverse), several isomorphisms between such parenthesizations
of four factors.  Specifically, the ``extreme left'' and ``extreme
right'' parenthesizations are connected by a product of three
$\alpha$'s:
\begin{multline*}
((A\otimes B)\otimes C)\otimes D
\overset{\alpha_{A,B,C}\otimes I_D}{\longrightarrow}
(A\otimes(B\otimes C))\otimes D
\overset{\alpha_{A,B\otimes C,D}}{\longrightarrow}\\
A\otimes ((B\otimes C)\otimes D)
\overset{I_A\otimes\alpha_{B,C,D}}{\longrightarrow}
A\otimes(B\otimes(C\otimes D)).
\end{multline*}
The same two parenthesizations are connected by a product of two other
$\alpha$'s:
\[
  ((A\otimes B)\otimes C)\otimes D
\overset{\alpha_{A\otimes B,C,D}}{\longrightarrow}
(A\otimes B)\otimes(C\otimes D)
\overset{\alpha_{A,B,C\otimes D}}{\longrightarrow}
A\otimes(B\otimes(C\otimes D)).
\]
One aspect of ``coherence'' is that these two transformations must
agree, so that there is a single, well-defined way of shifting the
parentheses from the left to the right. This requirement is often
called the \emph{pentagon condition}, because the diagram exhibiting
these two transformations together has the shape of a pentagon.  In
this connection, the first composition, involving three morphisms, is
sometimes called the ``long side'' of the pentagon, and the second
composition is the ``short side''.  In Figure~\ref{pentagon-fig}, the
short side is the top of the pentagon while the long side contains the
vertical sides and the bottom.

\begin{figure}
\xymatrix{
& (A\otimes B)\otimes(C\otimes D)
  \ar[1,1]^{\textstyle{\alpha}_{A,B,C\otimes D}}\\
((A\otimes B)\otimes C)\otimes D
  \ar[-1,1]^{\textstyle{\alpha}_{A\otimes B,C,D}}
  \ar[1,0]^{\textstyle{\alpha}_{A,B,C}\otimes \textstyle{I}_D} &
& A\otimes(B\otimes(C\otimes D)) \\
(A\otimes(B\otimes C))\otimes D
    \ar[0,2]_{\textstyle{\alpha}_{A,B\otimes C,D}} &
& A\otimes((B\otimes C)\otimes D)
  \ar[-1,0]^{\textstyle{I}_A\otimes\textstyle{\alpha}_{B,C,D}}
}
\caption{The pentagon condition}        \label{pentagon-fig}
\end{figure}

Another aspect of coherence is that two ways of simplifying
$(A\otimes1)\otimes B$ should agree, namely $\rho_A\otimes I_B$ and
\[
(A\otimes1)\otimes B
\overset{\alpha_{A,1,B}}{\longrightarrow}
A\otimes(1\otimes B)
\overset{I_A\otimes\lambda_B}{\longrightarrow}
A\otimes B.
\]

It is easy to think of other compositions of $\alpha$'s, $\lambda$'s,
and $\rho$'s that should agree, for example the many ways of
connecting different parenthesizations of five or more factors.
Fortunately, all of these requirements can be deduced from the two
that we have exhibited here.  This is Mac Lane's coherence theorem,
and we refer to Chapter VII of \cite{maclane} for its precise
statement, its proof, and additional information about monoidal
categories.

The pentagon condition will play a major role in the rest of this
paper, because the associativity isomorphism $\alpha$ is often
nontrivial and of considerable interest.  The unit isomorphisms
$\lambda$ and $\rho$, on the other hand, will play essentially no
role, because one can safely identify $1\otimes X$ and $X\otimes 1$
with $X$ and take $\lambda_X=\rho_X=I_X$ for all $X$.  From now on,
we will make these simplifying identifications.

The idea that $\otimes$ represents combining two anyons (or two
systems of anyons) into a single system suggests that this operation
should be commutative, i.e., that $X\otimes Y$ should be naturally
isomorphic to $Y\otimes X$.  The next axiom postulates the existence
of such an isomorphism, with good behavior in connection with the
associativity isomorphism $\alpha$.

\begin{ax}[\textbf{Braiding}]
The monoidal structure on \scr A is equipped with a \emph{braiding},
i.e., a natural isomorphism $\sigma_{X,Y}:X\otimes Y\to Y\otimes X$
subject to two requirements, first that the following two composite
isomorphisms be equal:
\[
(A\otimes B)\otimes C
\overset{\alpha_{A,B,C}}{\longrightarrow}
A\otimes(B\otimes C)
\overset{\sigma_{A,B\otimes C}}{\longrightarrow}
(B\otimes C)\otimes A
\overset{\alpha_{B,C,A}}{\longrightarrow}
B\otimes (C\otimes A)
\]
and
\[
(A\otimes B)\otimes C
\overset{\sigma_{A,B}\otimes I_C}{\longrightarrow}
(B\otimes A)\otimes C
\overset{\alpha_{B,A,C}}{\longrightarrow}
B\otimes(A\otimes C)
\overset{I_B\otimes \sigma_{A,C}}{\longrightarrow}
B\otimes (C\otimes A),
\]
and, second, the analogous equality with each $\sigma_{X,Y}$ replaced
with ${\sigma_{Y,X}}^{-1}$.
\end{ax}

\begin{figure}[h]
\[
\xymatrix{
& A\otimes(B\otimes C)
  \ar[0,1]^{\textstyle{\sigma}_{A,B\otimes C}}
& (B\otimes C)\otimes A
  \ar[1,1]^{\textstyle{\alpha}_{B,C,A}}
\\
(A\otimes B)\otimes C
  \ar[-1,1]^{\textstyle{\alpha}_{A,B,C}}
  \ar[1,1]_{\textstyle{\sigma}_{A,B}\otimes I_C}
&&& B\otimes (C\otimes A)
\\
& (B\otimes A)\otimes C
  \ar[0,1]_{\textstyle{\alpha}_{B,A,C}}
& B\otimes(A\otimes C)
  \ar[-1,1]_{\textstyle{I_B}\otimes \textstyle{\sigma}_{A,C}}
}
\]
\caption{The hexagon condition} \label{hexagon-fig}
\end{figure}

Recall, from Section~\ref{subsec:anyons}, that anyons inhabit
two-dimensional space and therefore, when two of them are
interchanged, it is necessary to keep track of how they move around
each other.  A clockwise rotation by $\pi$ around the midpoint between
them is not the same as, nor even deformable to, a counterclockwise
rotation.  So we should describe $\sigma_{X,Y}$ not merely as
switching $X$ with $Y$ but as doing so in a counterclockwise
direction. The choice of direction here is a matter of convention;
${\sigma_{Y,X}}^{-1}$ is then the clockwise rotation achieving the
same interchange.  Thus, we expect that, in general,
$\sigma_{X,Y}\neq{\sigma_{Y,X}}^{-1}$.  (If these two were always
equal, then we would have a \emph{symmetric monoidal} category rather
than a braided one.)

A useful picture, often used in connection with braiding, is to imagine
the factors in a $\otimes$-product as being lined up from left to
right. Then the counterclockwise interchange $\sigma_{X,Y}$ amounts to
moving $X$ from the left of $Y$ to the right of $Y$ by passing $X$ in
front of $Y$.  ${\sigma_{Y,X}}^{-1}$ also moves $X$ from
the left to the right of $Y$, but it does so by passing $X$ behind $Y$.

The equality of the two composite morphisms in the definition of
braiding is called the \emph{hexagon condition}
(Figure~\ref{hexagon-fig}). In terms of moving anyons around each
other, it expresses the fact that moving $A$ past $B\otimes C$ by
passing $A$ in front of $B\otimes C$ is equivalent to first passing
$A$ in front of $B$ and then passing $A$ in front of $C$.  The hexagon
condition for ${\sigma_{Y,X}}^{-1}$ has a similar pictorial
description with ``in front of'' replaced with ``behind''.

The last axiom in this subsection relates the multiplicative structure
discussed here with the additive structure from the preceding
subsection.

\begin{ax}[\textbf{Additive-Multiplicative}]
  \begin{lsnum}
    \item   The monoidal unit 1 is simple.
\item The product operation $\otimes$ is bilinear on morphisms.
  \end{lsnum}
\end{ax}

In more detail, item~(2) here means that the function
\[
\Hom(A,B)\times \Hom(C,D)\to\Hom(A\otimes C,B\otimes D)
\]
given by the functoriality of $\otimes$ is bilinear with respect to
the \bbb C-vector space structures of the Hom-sets.  It follows from
this, via results in \cite[Section~2.4]{freyd}, that $\otimes$
distributes over $\oplus$ on objects, i.e., that $X\otimes(Y\oplus
Z)$ is canonically isomorphic to $(X\otimes Y)\oplus(X\otimes Z)$.

\subsection{Duals, Twists, and Modularity}

In this subsection, we collect some additional axioms to complete the
definition of modular tensor categories.  These axioms will not play a
role in the computations we do later.  We list them for the sake of
completeness, but we make only a few comments about them and refer the
reader to \cite[Sections~4.3, 4.5, and 4.7]{panang} for more thorough
explanations.

\begin{ax}[\textbf{Antiparticles}]      \label{dual}
  For each object $X$ of \scr A, there is a \emph{dual} object $X^*$,
  and there are two morphisms $i_X:1\to X\otimes X^*$ and
  $e_X:X^*\otimes X\to 1$, such that the compositions
\[
X^*\overset{I_{A^*}\otimes i_X}{\longrightarrow}
X^*\otimes X\otimes X^*
\overset{e_X\otimes I_{X^*}}{\longrightarrow}X^*
\]
and
\[
X\overset{i_X\otimes I_X}{\longrightarrow}
X\otimes X^*\otimes X
\overset{I_X\otimes e_X}{\longrightarrow}X
\]
are equal to the identity morphisms $I_{X^*}$ and $I_X$,
respectively.  \looseness=-1 Furthermore, dualization commutes with $\otimes$ and
$\oplus$ and preserves 1 and 0.
\end{ax}

For the sake of readability, we have exhibited the compositions in
this axiom without the parentheses and associativity isomorphisms that
technically should be there.  We follow the same convention for
iterated $\otimes$ below.

The intention behind this axiom is that, if $X$ represents some
particle, then $X^*$ represents its antiparticle.  The morphism $i_X$
represents creation of a particle-antiparticle pair from the vacuum,
and $e_X$ represents annihilation of such a pair.

The operation of dualization becomes a contravariant functor from \scr
A to itself if one defines the dual $f^*$ of a morphism $f:X\to Y$ to
be the composite
\[
Y^*\overset{I_{Y^*}\otimes i_X}{\longrightarrow}
Y^*\otimes X\otimes X^*
\overset{I_{Y^*}\otimes f\otimes I_{X^*}}{\longrightarrow}
Y^*\otimes Y\otimes X^*
\overset{e_Y\otimes I_{X^*}}{\longrightarrow}
X^*.
\]

\begin{ax}[\textbf{Rotations}]
There is a natural isomorphism $\delta$ with components $\delta_X:X\to
X^{**}$  respecting the monoidal structure and duality in the sense
that
\[
\delta_1=I_1,\quad \delta_{X\otimes Y}=\delta_X\otimes\delta_Y,
\quad\text{and}\quad \delta_{X^*}=({\delta_X}^*)^{-1}.
\]
\end{ax}

By combining these $\delta$ isomorphisms with the morphisms $i$ and
$e$ from duality, one can obtain isomorphisms $X\to X$ that represent
twisting an anyon by $2\pi$; see \cite[Section~4.5]{panang} for
details.

Monoidal categories satisfying the ``Antiparticles'' axiom are called
\emph{rigid}, and those that also satisfy the ``Rotations'' axiom are
called \emph{ribbon} categories.

\begin{ax}[\textbf{Modularity}]
For any two simple objects $X$ and $Y$, let $s_{X,Y}:1\to1$ be the
morphism
\begin{multline*}
  1=1\otimes 1
\overset{i_X\otimes i_Y}{\longrightarrow}
X\otimes X^*\otimes Y\otimes Y^*
\overset{I_X\otimes\sigma_{X^*,Y}\otimes I_Y}{\longrightarrow}
X\otimes Y\otimes X^*\otimes Y^*\\
\overset{I_X\otimes\sigma_{Y,X^*}\otimes I_Y}{\longrightarrow}
X\otimes X^*\otimes Y\otimes Y^*
\overset{\delta_X\otimes I_{X^*}\otimes \delta_Y\otimes
  I_{Y^*}}{\longrightarrow}
X^{**}\otimes X^*\otimes Y^{**}\otimes Y^*\\
\overset{e_{X^*}\otimes e_{Y^*}}{\longrightarrow}
1\otimes 1=1.
\end{multline*}
Since $\Hom(1,1)=\bbb C$, these morphisms $s_{X,Y}$ constitute a
matrix of complex numbers, with rows and columns indexed by the
isomorphism classes of simple objects.  This matrix is required to be
invertible.
\end{ax}

Notice that, if \scr A were not merely braided but symmetric, then the
$\sigma$'s and the $\sigma^{-1}$'s in this composite would cancel out,
and we would have $s_{X,Y}=t_Xt_Y$ where $t_X$ is the composite
\[
1\overset{i_X}{\longrightarrow}X\otimes X^*
\overset{\delta_X\otimes I_{X^*}}{\longrightarrow}X^{**}\otimes X^*
\overset{e_{X^*}}{\longrightarrow}1,
\]
and similarly for $t_Y$.  Thus, the matrix $s$ described in the modularity
axiom would be the product of a column vector by a row vector (in this
order).  Such a matrix has rank at most 1. By requiring this matrix to be
invertible, the axiom says that, as far as the rank of this matrix is
concerned, the braiding is as far as possible from being symmetric.

\section{Yoneda simplification}         \label{yoneda}

In this section, we point out a simplification of the additive
structure of \scr A, based on Yoneda's Lemma.  That lemma (see
\cite[Section~III.2]{maclane}) says roughly that an object in any
category is determined, up to isomorphism, by the morphisms into it.
More precisely, any category \scr C is equivalent to a full
subcategory of the category $\hat{\scr C}$ of contravariant functors
from \scr C to the category of sets.\footnote{There are set-theoretic
  issues if \scr C is a proper class rather than a set, but these
  issues need not concern us here.  The finiteness conditions imposed
  on our anyon category \scr A ensure that it is equivalent to a
  small, i.e., set-sized, category.}  Under this equivalence, any
object $X$ of \scr C corresponds to the functor $\Hom(-,X)$, i.e., the
functor sending each object $U$ of \scr C to the set of morphisms
$U\to X$ and sending each morphism $f:U\to V$ to the operation
$\Hom(V,X)\to\Hom(U,X)$ of composition with $f$.

In the case of our category \scr A, we can greatly simplify $\hat{\scr
  A}$ while still maintaining the Yoneda equivalence.  In the first
place, since every object $U$ of \scr A is a finite sum, and thus in
particular a coproduct, of simple objects, $U=\bigoplus_{j\in F}S_j$,
morphisms $U\to X$ amount to $F$-indexed families of morphisms $S_j\to
X$.  More precisely, any $f:U\to X$ is determined by the composite
morphisms $f\circ u_j:S_j\to X$, and, conversely, any family of
morphisms $g_j:S_j\to X$ arises in this way from a unique morphism
$U\to X$.  Thus, \scr A is equivalent to a full subcategory of the
category $\hat{\scr S}$ of set-valued functors on the category \scr S
of simple objects in \scr A.

Up to equivalence, we need not use all the simple objects; it suffices
to have at least one representative from each isomorphism class of
simple objects.  So we can replace the \scr S of the preceding
paragraph by a skeleton of it, i.e., a full subcategory $\scr S_0$
consisting of just one representative per isomorphism class.

The structure of this new, skeletal $\scr S_0$ admits, thanks to the
finiteness axiom and Proposition~\ref{schur} the following
description. There are finitely many objects.  The morphisms from any
object to itself form a copy of \bbb C.  If $U$ and $V$ are distinct
objects, then the only morphism from $U$ to $V$ is zero.

As a result, the Yoneda embedding, simplified as above, sends each
object $X$ of \scr A to a finite family of vector spaces, indexed by
the simple objects $U$ in $\scr S_0$, namely the vector spaces
$\Hom(U,X)$.  Furthermore, the morphisms $X\to Y$ in \scr A are given
by \emph{arbitrary} families of linear maps
$g_U:\Hom(U,X)\to\Hom(U,Y)$ between corresponding vector spaces.  The
reason for ``arbitrary'' is that, because of the paucity of morphisms
in $\scr S_0$, all such families automatically satisfy the
commutativity conditions required in order to be natural
transformations and thus to be morphisms in the functor category
$\hat{\scr S}_0$.

Summarizing, we have that, up to equivalence of categories, \scr A can
be described as the category whose objects (resp.\ morphisms) are
families of finite-dimensional vector spaces (resp.\ linear maps),
indexed by the objects of $\scr S_0$.  Furthermore, it is easy to
check that sums in \scr A are given, via this equivalence, by direct
sums of vector spaces.

In other words, the additive structure of \scr A is trivial.  The
interesting structure is the monoidal structure, and this can be quite
complicated.  In particular, the associativity isomorphisms $\alpha$
and the braiding isomorphisms $\sigma$, though given (like any
morphisms) by linear maps, need not have a particularly simple
structure.

The analysis of the multiplicative structure of \scr A can be
facilitated by taking advantage of the semisimplicity of \scr A and
the fact that $\otimes$ distributes over $\oplus$.  If we know how
$\otimes$ acts on simple objects, distributivity determines how it
acts on sums of simple objects, and, by semisimplicity, those are all
the objects.  Moreover, because the associativity and braiding
isomorphisms are natural, and thus in particular commute with the
injection and projection morphisms of sums, the behavior of these
isomorphisms on arbitrary objects is determined by their behavior on
simple objects.  Better yet, the pentagon and hexagon conditions will
be satisfied in general as soon as they are satisfied for simple
objects.

Thus, the additive and multiplicative structure of \scr A can be
completely described by giving
\begin{lsnum}
  \item a complete list of non-isomorphic simple objects (including
    the unit or vacuum 1),
\item for each pair of objects in this list, their $\otimes$-product,
  expressed as a sum of objects from the list,
\item the associativity isomorphisms $\alpha_{X,Y,Z}$ for all $X,Y,Z$
  in the list, and
\item the braiding isomorphisms $\sigma_{X,Y}$ for all $X,Y$ in the list,
\end{lsnum}
subject to the pentagon and hexagon conditions.

We shall not be concerned here with duality and ribbon structure, but
it could also be reduced to a consideration of the simple cases.

Often, items~(1) and (2) here determine or at least greatly constrain
items~(3) and (4) via the pentagon and hexagon conditions.  One such
situation is the subject of the next section.  Other examples, both of
strong constraints on (3) and (4) and of weak constraints can be found
in \cite{bonder}.

\section{Fibonacci anyons}              \label{fib}

\subsection{Definition and Additive Structure}

In this section, we consider the special case of \emph{Fibonacci
  anyons}.  These are defined by specifying the category \scr A as
follows.  There are just two simple objects, 1 (the vacuum, the unit
for $\otimes$) and $\tau$.  Each is its own dual.  (Recall that
Axiom~\ref{dual} requires each object to have a dual; dualization is
additive, so we need only specify the duals of the simple objects.)
The monoidal structure is given by $\tau\otimes\tau=1\oplus\tau$ (plus
the fact that 1 is the unit, so $1\otimes\tau=\tau\otimes1=\tau$ and
$1\otimes 1=1$).

The terminology ``Fibonacci anyon'' comes from the fact, easily
verified using the distributivity of $\otimes$ over $\oplus$, that
iteration of $\otimes$ gives $\tau^{\otimes n}=f_{n-1}\cdot1\oplus
f_n\cdot\tau$, where the $f$'s are the Fibonacci numbers defined by
the recursion $f_{-1}=1$, $f_0=0$, and $f_{n+1}=f_n+f_{n-1}$.  Here
and below, we use the notation $k\cdot S$ to mean the sum of $k$
copies of the object $S$ of \scr A.  (The notation makes sense for
arbitrary objects $S$, but we shall need it only for simple $S$.)

As explained in Section~\ref{yoneda}, we can identify the category
\scr A with the category of pairs $(V_1,V_\tau)$ of finite-dimensional
complex vector spaces.  Explicitly, an object $X$ is identified with
the pair $(\Hom(1,X),\Hom(\tau,X))$.  In particular, the unit 1 in
\scr A is identified with $(\bbb C,0)$, and $\tau$ is identified with
$(0,\bbb C)$.  This identification respects the additive structure:
$\oplus$ in \scr A corresponds to componentwise direct sum of pairs of
vector spaces.

\subsection{Tensor Products}

The multiplicative structure of \scr A, on the other hand, is quite
far from componentwise tensor product of vector spaces, as the latter
would make $\tau\otimes\tau=\tau$ (because $\bbb C\otimes\bbb
C\cong\bbb C)$. Our goal in the rest of this paper
is to determine the multiplicative structure in terms of pairs of
vector spaces.

The equation $\tau^n=f_{n-1}\cdot1\oplus f_n\cdot\tau$ mentioned above
already determines that structure as far as the objects are concerned,
but there remains much to be said about the morphisms.

A morphism from one pair of vector spaces $(V_1,V_\tau)$ to another
such pair $(W_1,W_\tau)$ is a pair of linear transformations
$(m_1:V_1\to W_1,m_\tau:V_\tau\to W_\tau)$.  We can think of it as a
pair of matrices, provided we fix bases for all the vector spaces
involved here.

The choice of bases involves considerable arbitrariness, but there is
a (somewhat) helpful guiding principle, namely that, if we have
already chosen bases for two vector spaces, then the union of those
bases serves naturally as a basis for the direct sum of those vector spaces.
Some caution is needed, though, because the same vector space can
arise as a direct sum in several ways and can thus have several
equally natural bases.  Indeed, much of our work below will be finding
the transformations that relate such bases.

The guiding principle tells us nothing about choosing bases for the
one-dimensional spaces $V_1$ and $V_\tau$ in the pairs $1=(V_1,0)$ and
$\tau=(0,V_\tau)$.  There isn't even any non-zero morphism between
these simple objects to suggest a correlation between the choice of
bases.  Nor do we get canonical bases here by evaluating compound
expressions that fuse to $\tau$ or to $1$ or to a sum of these.  So we
might as well identify these one-dimensional spaces with \bbb C and
use the number 1 as the basis vector in both of them.

Then $\tau\otimes\tau=1\oplus\tau=(\bbb C,\bbb C)$ already has a basis
for each of the two vector spaces.  Let us turn to the triple product
\[
\tau\otimes(\tau\otimes\tau)=\tau\otimes(1\oplus\tau)=
(\tau\otimes1)\oplus(\tau\otimes\tau)=\tau\oplus(1\oplus\tau)=
1\cdot1\oplus2\cdot\tau.
\]
As a pair of vector spaces, it is isomorphic to $(\bbb C,\bbb C^2)$,
but we have some additional information about it, namely that it was
obtained as the sum of $\tau\otimes1=\tau$ and
$\tau\otimes\tau=1\oplus\tau$. Our guiding principle thus suggests
choosing a basis in $\bbb C^2$ that respects this sum decomposition.
That is, one of the basis vectors in $\bbb C^2$ should come from the
first $\tau$ and the other should come from the second summand,
$1\oplus\tau$.

Consider, however, the analogous computation with the other way of
parenthesizing the triple product:
\[
(\tau\otimes\tau)\otimes\tau=(1\oplus\tau)\otimes\tau=
(1\otimes\tau)\oplus(\tau\otimes\tau)=\tau\oplus(1\oplus\tau)=
1\cdot1\oplus2\cdot\tau.
\]
It also leads to the pair of vector spaces $(\bbb C,\bbb C^2)$, and it
also provides a suggestion for a basis of $\bbb C^2$.  There is,
however, no guarantee that this suggestion agrees with the one in the
preceding paragraph. We shall see below that the two suggestions are
actually guaranteed to \emph{disagree}.  We have two bases for $\bbb
C^2$, and there will be a non-trivial matrix transforming the one into
the other.  We shall find that this matrix is almost uniquely
determined.

There could, a priori, have also been two different natural bases
for the first component \bbb C in $\tau^{\otimes 3}$, although we shall see that, in this particular situation, they coincide.

These basis transformation matrices, relating the bases that arise
from $\tau\otimes(\tau\otimes\tau)$ and from
$(\tau\otimes\tau)\otimes\tau$, amount to the associativity
isomorphism $\alpha_{\tau,\tau,\tau}$ in the definition of the
monoidal category \scr A.

Recall from Section~\ref{yoneda} that all the associativity
isomorphisms of \scr A are determined by those with simple objects as
subscripts.  One of these is the $\alpha_{\tau,\tau,\tau}$ mentioned
just above; the others involve one or more 1's in the subscript.
Fortunately, all those others are identity maps, thanks to the
identification of $1\otimes X$ and $X\otimes1$ with $X$.  So the
entire associativity structure of \scr A comes down to two matrices, a
$2\times 2$ matrix relating the two bases for $\bbb C^2$ and a number
(a $1\times 1$ matrix) relating the two bases for \bbb C.  These
matrices are subject to the constraint given by the pentagon
condition (Figure~\ref{pentagon-fig}).
Below, we shall calculate that constraint
explicitly.  It will almost uniquely determine $\alpha$.

We shall also calculate the constraint imposed by the hexagon
condition on the braiding isomorphisms $\sigma$
(Figure~\ref{hexagon-fig}).  Again, the only component that needs to
be computed is $\sigma_{\tau,\tau}$.  The components where at least
one subscript is 1 are trivial, and the components with non-simple
objects as subscripts reduce, by distributivity, to ones with simple
subscripts.

\subsection{Notation for Basis Vectors}
In order to compute the isomorphisms $\alpha_{\tau,\tau,\tau}$ and
$\sigma_{\tau,\tau}$ for Fibonacci anyons, we shall view them as
matrices, using suitable bases for the relevant vector spaces, and we
shall calculate the constraints imposed on those matrices by the
pentagon and hexagon conditions.  We begin by setting up a convenient
notation for those bases.

The domains and codomains of the morphisms under consideration are
obtained from $\tau$ and 1 by iterated $\otimes$.  We must, of
course, be careful about the parenthesization of such
$\otimes$-products because, as we saw above, different
parenthesizations can lead to different bases; indeed,
$\alpha_{\tau,\tau,\tau}$ contains exactly the information about how
two such bases are related.

In general, given a parenthesized $\otimes$-product of $\tau$'s and
1's, we can use the defining equations for Fibonacci anyons,
particularly $\tau\otimes\tau=1\oplus\tau$, and the distributivity of
$\otimes$ over $\oplus$, to convert the given product into a sum of
$\tau$'s and 1's.  Each summand in that sum arises from the original
product as a result of certain choices of 1 or $\tau$ when expanding
some occurrences of $\tau\otimes\tau$.

For example, in the equation
\[
\tau\otimes(\tau\otimes\tau)=\tau\otimes(1\oplus\tau)=
(\tau\otimes1)\oplus(\tau\otimes\tau)=\tau\oplus(1\oplus\tau)=
1\cdot1\oplus2\cdot\tau
\]
considered above, the summand 1 at the right end of the equation arose
from the $\tau\otimes(\tau\otimes\tau)$ at the left end by first
choosing the summand $\tau$ in the evaluation of $(\tau\otimes\tau)$
at the first step in the equation, and then, after applying the
distributive law at the second step, choosing the summand 1 in the
evaluation of $\tau\otimes\tau$ at the third step.  These choices can
be visualized as the tree
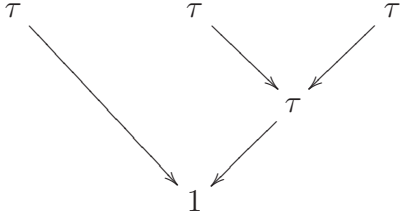
\[
\xymatrix{
{\tau}\ar[2,2] &&{\tau} \ar[dr] && {\tau} \ar[dl] \\
&&& {\tau} \ar[dl]          \\
&& 1
}
\]
or, in a more compressed notation,
\[
\one\tau{\two \tau\tau}.
\]
Here the three $\tau$'s and the parentheses describe the
$\otimes$-product $\tau\otimes(\tau\otimes\tau)$ that we began with,
and the symbols under the dots indicate the choice of summand at each
step.  The inner $\underset{\tau}{\cdot}$ indicates that, from the
evaluation of the inner $\tau\otimes\tau=1\oplus\tau$, we chose the
$\tau$ summand.  After applying distributivity, that leads us to
$\tau\otimes\tau$, from which, as indicated by the outer
$\underset{1}{\cdot}$, we chose the summand 1.

The other possible choices during the same evaluation
would be written
\[
\two \tau{\two \tau\tau} \quad\text{and}\quad
\two \tau{\one \tau\tau}
\]
and depicted by the trees
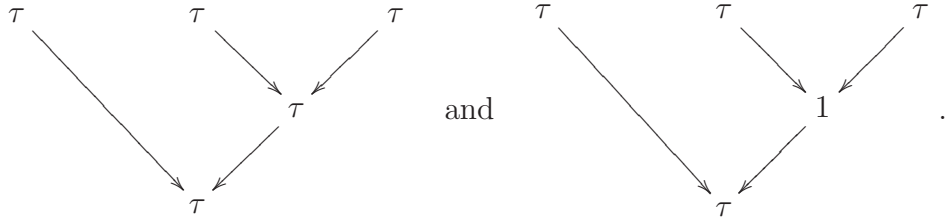
\[
\begin{minipage}{.5\textwidth}\centering
\xymatrix{
{\tau}\ar[2,2] &&{\tau} \ar[dr] && {\tau} \ar[dl] \\
&&& {\tau} \ar[dl]          \\
&& {\tau}
}
\end{minipage}
\quad\text{and}\quad
\begin{minipage}{.5\textwidth}\centering
\xymatrix{
{\tau}\ar[2,2] &&{\tau} \ar[dr] && {\tau} \ar[dl] \\
&&& 1\ar[dl]          \\
&& {\tau}
}
\end{minipage}.
\]
The first of these indicates that, as before, we chose the $\tau$
summand when evaluating the inner $\otimes$, obtaining, when
distributivity is applied, the summand $\tau\otimes\tau=1\oplus\tau$,
but then we chose the $\tau$ rather than the 1.  The second indicates
that, when evaluating the inner $\tau\otimes\tau$, we chose the
summand 1, so that, after applying distributivity, we got
$\tau\otimes1$.  Here, there is no choice remaining to be made;
$\tau\otimes1$ is simply $\tau$.  Nevertheless, we write $\tau$ under
the outer dot and at the root of the tree, to make it obvious that the
final result here is $\tau$.

In what follows, we shall systematically use the compressed notation,
but the reader can easily draw the tree diagrams.  Indeed, these
diagrams are just the parse trees of the compressed notations.  The
trees can also be viewed as a sort of Feynman diagrams, depicting how
the anyons at the leaves of the tree fuse on their way to the root.

In our notation, we write a product of $\tau$'s or 1's
with $\tau$'s or 1's also under the dots, to represent
specific summands (1 or $\tau$) in the fully distributed expansion of
a $\otimes$-product of $\tau$'s and 1's.  To evaluate $\one XY$, first
evaluate $X$ and $Y$; then apply $\otimes$ to them; and then take the
1 summand in the result. To evaluate $\two XY$ do the same except that
you take the $\tau$ summand in the result.  These notations will never
be used in situations where they would be meaningless because the
required summand is not present in the result; that is, we never write
$\one XY$ when one of $X,Y$ evaluates to 1 and the other to $\tau$, for
then $\otimes$ yields only $\tau$; and we never write $\two XY$ when both
of $X,Y$ evaluate to 1. As in one of the examples above, we include
the subscript under the dot even when that subscript is forced
because one of the factors evaluates to 1.

Notice that our notation provides symbols, like the three examples
above, that denote not only an object 1 or $\tau$ (which can be read
off by just looking under the outermost dot in the notation) but also
a particular occurrence of that $1=(\bbb C,0)$ or $\tau=(0,\bbb C)$ as
a subspace (direct summand) of a specific $\otimes$-product, namely
the product with the same factors and the same parentheses as in our
notation.

In other words, if we are given a parenthesized $\otimes$-product of
1's and $\tau$'s, representing the pair of vector spaces
$(V_1,V_\tau)$, then by replacing each $\otimes$ by either
$\underset{1}{\cdot}$ or $\underset{\tau}{\cdot}$, we obtain (either a
meaningless expression because some required summand is absent or) a
notation for a subspace of $V_1$ or $V_\tau$.  It denotes a subspace
of $V_1$ (resp.\ $V_\tau$) just in case the outermost $\otimes$ was
replaced by $\underset{1}{\cdot}$ (resp.\ $\underset{\tau}{\cdot}$).

Our notation provides names for certain summands $1=(\bbb C,0)$ or
$\tau=(0,\bbb C)$ of certain objects $(V_1,V_\tau)$ of the Fibonacci
category \scr A.  We shall also use the same notation for the resulting
basis vectors.  That is, once we have a copy of, say, $(\bbb C,0)$ in
$(V_1,V_\tau)$, the number 1 in \bbb C corresponds to some vector in
$V_1$, and we shall use the same notation for this vector as for the
summand.  The same goes for the case of copies of $(0,\bbb C)$ in
$(V_1,V_\tau)$; they provide vectors in $V_\tau$.

Notice that, if we begin with some parenthesized $\otimes$-product of
1's and $\tau$'s, with value $(V_1,V_\tau)$ in \scr A, and if we form
all possible (meaningful) notations by replacing $\otimes$ by
$\underset{1}{\cdot}$ or $\underset{\tau}{\cdot}$, then the resulting
vectors, as described in the preceding paragraph, constitute bases for
the vector spaces $V_1$ and $V_\tau$.  This observation is just a
restatement of the fact that the original parenthesized
$\otimes$-product is the direct sum of all the simple objects
obtainable by making the choices indicated by the subscripts in our
notation.

\subsection{Associativity}
Now that we have a general notation system for the basis vectors in
parenthesized $\otimes$-products, we turn to the specific cases
involved in associativity and the pentagon condition.

The unique ``interesting'' component of associativity,
$\alpha_{\tau,\tau,\tau}$, which we sometimes abbreviate as simply
$\alpha$, is an isomorphism from $(\tau\otimes\tau)\otimes\tau$ to
$\tau\otimes(\tau\otimes\tau)$, both of which are, as pairs of vector
spaces, a 1-dimensional $V_1$ and a 2-dimensional $V_\tau$.  The first
parenthesization gives a basis vector
\[
\one {\two \tau\tau}\tau \quad\text{for }V_1
\]
and two basis vectors
\[
\two {\one \tau\tau}\tau \quad\text{and}\quad
\two {\two \tau\tau}\tau \quad\text{for }V_\tau.
\]
The second parenthesization similarly gives a basis vector
\[
\one \tau{\two \tau\tau} \quad\text{for }V_1
\]
and two basis vectors
\[
\two \tau{\one \tau\tau} \quad\text{and}\quad
\two \tau{\two \tau\tau} \quad\text{for }V_\tau.
\]
Our task is to compute the transformation $\alpha$ between these
bases\footnote{We have chosen to regard $V_1$ and $V_\tau$ as each
  being a single space, independent of the parenthesization.  The
  different parenthesizations give (possibly) different bases for
  these spaces.  An alternative view is that each parenthesization
  gives its own $V_1$ and $V_\tau$, isomorphic to \bbb C and $\bbb
  C^2$ respectively, with their standard bases, while $\alpha$ gives
  an isomorphism between the two $V_1$'s and an isomorphism between
  the two $V_\tau$'s.  The two viewpoints are easily intertranslatable
  and the computations that follow would be the same in either
  picture.}.
This $\alpha$ has two components, the first relating two
bases of the one-dimensional space $V_1$ and the second relating two
bases of the two-dimensional space $V_\tau$.  These are given,
respectively, by a non-zero number $p$ such that
\[
\one {\two \tau\tau}\tau = p\one \tau{\two \tau\tau}
\]
and a non-singular matrix $
\begin{pmatrix}
  q&r\\s&t
\end{pmatrix}
$ such that
\begin{align*}
  \two {\one \tau\tau}\tau&=q\two \tau{\one \tau\tau}+r\two \tau{\two
    \tau\tau}\\
  \two {\two \tau\tau}\tau&=s\two \tau{\one \tau\tau}+t\two \tau{\two
    \tau\tau}.
\end{align*}
Here ``non-zero'' for $p$ and ``non-singular'' for the matrix embody
the requirement that $\alpha$ is an isomorphism.

We shall now investigate the constraints imposed on $p,q,r,s,t$ by the
pentagon condition.

\medskip
\xymatrix{
& (\tau\otimes \tau)\otimes(\tau\otimes \tau)
  \ar[1,1]^{\textstyle{\alpha}_{\tau,\tau,\tau\otimes \tau}}\\
((\tau\otimes \tau)\otimes \tau)\otimes \tau
  \ar[-1,1]^{\textstyle{\alpha}_{\tau\otimes \tau,\tau,\tau}}
  \ar[1,0]_{\textstyle{\alpha}_{\tau,\tau,\tau}\otimes \textstyle{I}_\tau} &
& \tau\otimes(\tau\otimes(\tau\otimes \tau)) \\
(\tau\otimes(\tau\otimes \tau))\otimes \tau
    \ar[0,2]_{\textstyle{\alpha}_{\tau,\tau\otimes \tau,\tau}} &
& \tau\otimes((\tau\otimes \tau)\otimes \tau)
  \ar[-1,0]_{\textstyle{I}_\tau\otimes\textstyle{\alpha}_{\tau,\tau,\tau}}
}
\medskip

\noindent That condition involves the $\otimes$-product of four
$\tau$'s, parenthesized in five ways, and we shall need to consider
the natural bases for all five parenthesizations.  Since
$\tau^{\otimes4}=(\bbb C^2,\bbb C^3)$, each parenthesization will give
two vectors as a basis for the 1 component and three as a basis for
the $\tau$ component.  We begin by considering the $\tau$ components,
whose bases are displayed below.  (There is no significance to the
chosen ordering of the five bases, nor the ordering of the three
vectors within each basis.)
\[
\begin{matrix}
\two {\one {\two \tau\tau}\tau}\tau&
\two {\two {\one \tau\tau}\tau}\tau&
\two {\two {\two \tau\tau}\tau}\tau\\
\two {\one \tau\tau}{\two \tau\tau}&
\two {\two \tau\tau}{\one \tau\tau}&
\two {\two \tau\tau}{\two \tau\tau}\\
\two {\one \tau{\two \tau\tau}}\tau&
\two {\two \tau{\one \tau\tau}}\tau&
\two {\two \tau{\two \tau\tau}}\tau\\
\two \tau{\one {\two \tau\tau}\tau}&
\two \tau{\two {\one \tau\tau}\tau}&
\two \tau{\two {\two \tau\tau}\tau}\\
\two \tau{\one \tau{\two \tau\tau}}&
\two \tau{\two \tau{\one \tau\tau}}&
\two \tau{\two \tau{\two \tau\tau}}
\end{matrix}
\]
Each row in this picture is a basis for the 3-dimensional $V_\tau$;
specifically, it is the basis arising from the same parenthesization
of $\tau\otimes\tau\otimes\tau\otimes\tau$ as the parenthesization in
our notation.

When writing transformation matrices between these bases, we must
regard each basis as given in a specific order, because rows of a
matrix come in an order.  We (arbitrarily) choose the orders in which
the bases are displayed above.

The five isomorphisms that appear in the pentagon condition amount to
five transformations between these bases.
Let us consider these one
at a time, beginning with the one connecting the first two bases in
the table.  Here we are dealing with the isomorphism
\[
\alpha_{\tau\otimes\tau,\tau,\tau}:
(((\tau\otimes\tau)\otimes)\tau\otimes\tau)\to
((\tau\otimes\tau)\otimes(\tau\otimes\tau)).
\]
The first subscript of this $\alpha$, namely $\tau\otimes\tau$, can be
decomposed as the sum $1\oplus\tau$, and the naturality of $\alpha$
then implies that $\alpha_{\tau\otimes\tau,\tau,\tau}$ is the direct
sum of $\alpha_{1,\tau,\tau}$ and $\alpha_{\tau,\tau,\tau}$.  The
first of these two summands is the identity, like all associativity
isomorphisms where one of the three factors is 1.  The second summand
is given by our matrix $
\begin{pmatrix}
  q&r\\s&t
\end{pmatrix}$.  As a result, we find that the transformation
$\alpha_{\tau\otimes\tau,\tau,\tau}$ connecting the first two bases in
our list is (taking into account the order in which the basis vectors
are listed)
\[
\alpha_{\tau\otimes\tau,\tau,\tau}=
\begin{pmatrix}
  0&q&r\\1&0&0\\0&s&t
\end{pmatrix}.
\]
In this matrix, the 1 in the (2,1) position and the two zeros in its
row arise from the fact that the identity map $\alpha_{1,\tau,\tau}$
sends the second vector in our first basis to the first vector in the
second basis.  Had we listed $\two{\two{\one\tau\tau}\tau}\tau$ first
rather than second in our first basis, the matrix for
$\alpha_{\tau\otimes\tau,\tau,\tau}$ would have been a block diagonal
matrix with 1 in the upper left corner.

An exactly analogous computation gives the isomorphism between the
second and the last bases in our list:
\[
\alpha_{\tau,\tau,\tau\otimes\tau}=
\begin{pmatrix}
  q&0&r\\0&1&0\\s&0&t
\end{pmatrix}
\]

Multipying these two matrices, we get the transformation from the
first basis (parenthesized to the left) to the last (parenthesized to
the right) that corresponds to the ``short'' side of the pentagon (two
morphisms, across the top of the diagram).  This product is
\[
\begin{pmatrix}
  rs&q&rt\\q&0&r\\st&s&t^2
\end{pmatrix}.
\]

Turning to the long side of the pentagon (three morphisms), we find
that the middle one, corresponding to rows 3 and 4 in our list of
bases and to the bottom of the diagram, is quite analogous to the two
that we have already computed.  It is
\[
\alpha_{\tau,\tau\otimes\tau,\tau}=
\begin{pmatrix}
  q&0&r\\0&1&0\\s&0&t
\end{pmatrix}.
\]

The remaining two isomorphisms for the long side of the pentagon (the
vertical arrows in the diagram) are a bit different, as they involve
$\alpha$'s on three of the four factors and an identity map on the
remaining factor.  Let us consider $\alpha_{\tau,\tau,\tau}\otimes
I_\tau$, which connects the first basis in our list to the third.  In
effect, this ignores the rightmost factor and acts like $\alpha$ on
the first three factors.  In other words, it is given by the same
matrix as the transformation from the basis
\[
\one {\two \tau\tau}\tau \qquad \two {\one \tau\tau}\tau \qquad
\two {\two \tau\tau}\tau
\]
to the basis
\[
\one \tau{\two \tau\tau} \qquad \two \tau{\one \tau\tau} \qquad
\two \tau{\two \tau\tau}.
\]
Notice that, in each of these bases the first element is in the $V_1$
component, so that component of $\alpha$, namely $p$, enters the
picture.  Indeed, the matrix connecting these bases is
\[
\alpha_{\tau,\tau,\tau}\otimes I_\tau=
\begin{pmatrix}
  p&0&0\\0&q&r\\0&s&t
\end{pmatrix}.
\]

Similarly, the remaining isomorphism on the long side of the pentagon
is also
\[
I_\tau\otimes\alpha_{\tau,\tau,\tau}=
\begin{pmatrix}
  p&0&0\\0&q&r\\0&s&t
\end{pmatrix}.
\]

Multiplying the three matrices for the long side of the pentagon, and
equating, as the pentagon condition requires, the resulting product to
the product that we obtained for the short side of the pentagon, we
have
\[
\begin{pmatrix}
  p^2q&prs&prt\\prs&q^2+rst&qr+rt^2\\pst&qs+st^2&rs+t^3
\end{pmatrix}=
\begin{pmatrix}
  rs&q&rt\\q&0&r\\st&s&t^2
\end{pmatrix}.
\]
This is the $V_\tau$ part of the pentagon condition.  Before turning
to the $V_1$ part, let us extract as much information as possible from
the matrix equation that we have just derived.

Suppose, toward a contradiction, that $p\neq1$.  Then the (1,3) and
(3,1) components of our matrix equation give $rt=st=0$, so either
$r=s=0$ or $t=0$.  If $r=s=0$, then the (1,2) component of the matrix
equation gives that $q=0$ also, but this contradicts the fact that $
\begin{pmatrix}
  q&r\\s&t
\end{pmatrix}$ is non-singular. There remains the case that $t=0$.
Then the (2,2) component says $q=0$, the (2,3) component says $r=0$,
and we again contradict the non-singularity of $
\begin{pmatrix}
  q&r\\s&t
\end{pmatrix}$.  So we have contradictions in all cases if $p\neq1$.

So $p=1$.  Now the (1,1) entry of the matrix equation gives $q=rs$.
Substituting that into the (2,2) component, we get $q(q+t)=0$, so
either $q=0$ or $q=-t$.  The first of these options leads, via the
(1,2) entry, to $rs=0$ and thus to a contradiction to non-singularity,
as before.  Therefore $q=-t$.

From the (2,3) and (3,2) entries, we get that $(q+t^2)r=r$ and
$(q+t^2)s=s$.  We cannot have both $r=0$ and $s=0$, as that would give
$q=0$ in the (1,2) entry and contradict non-singularity.  So we must
have $q+t^2=1$.  In view of $q=-t$, this means $q^2+q-1=0$ and
therefore
\[
q=-t=\frac{-1\pm\sqrt5}2.
\]
This evaluation of $q$ and $t$, together with the earlier results
\[
p=1\qquad\text{and}\qquad rs=q,
\]
satisfy, as one easily checks, the entire matrix equation above.  The
least trivial item to check is the (3,3) entry, $rs+t^3=t^2$, which,
in view of the equations above, becomes $q-q^3=q^2$, i.e.,
$0=q(q^2+q-1)$, and this is true because $q$ was obtained as a
solution of $q^2+q-1=0$.

All of the preceding calculation was based on the $V_\tau$ component
of $\tau^{\otimes 4}$; we still have the $V_1$ component of the
pentagon equation to work out.  Again, we have a list of five bases,
now for a 2-dimensional space, as follows.
\[
\begin{matrix}
  \one {\two {\one \tau\tau}\tau}\tau&
\one {\two {\two \tau\tau}\tau}\tau\\
\one {\one \tau\tau}{\one \tau\tau}&
\one {\two \tau\tau}{\two \tau\tau}\\
\one {\two \tau{\one \tau\tau}}\tau&
\one {\two \tau{\two \tau\tau}}\tau\\
\one \tau{\two {\one \tau\tau}\tau}&
\one \tau{\two {\two \tau\tau}\tau}\\
\one \tau{\two \tau{\one \tau\tau}}&
\one \tau{\two \tau{\two \tau\tau}}
\end{matrix}
\]

Computations analogous to (but shorter than) the earlier ones give,
for the short side of the pentagon,
\[
\alpha_{\tau\otimes\tau,\tau,\tau}=
\begin{pmatrix}
  1&0\\0&p
\end{pmatrix}
\text{ and }
\alpha_{\tau,\tau,\tau\otimes\tau}=
\begin{pmatrix}
  1&0\\0&p
\end{pmatrix}.
\]
So the product for the short side is simply $
\begin{pmatrix}
  1&0\\0&p^2
\end{pmatrix}$.
For the long side, we get
\[
\alpha_{\tau,\tau\otimes\tau,\tau}=
\begin{pmatrix}
  1&0\\0&p
\end{pmatrix}
\]
and
\[
\alpha_{\tau,\tau,\tau}\otimes
I_\tau=I_\tau\otimes\alpha_{\tau,\tau,\tau} =
\begin{pmatrix}
  q&r\\s&t
\end{pmatrix}.
\]
Equating the product of the long side and the product of the short
side, we get
\[
\begin{pmatrix}
  1&0\\0&p^2
\end{pmatrix}=
\begin{pmatrix}
  q^2+prs&qr+ptr\\qs+pts&rs+pt^2
\end{pmatrix}.
\]

This matrix equation is automatically satisfied because of the
equations that we had already derived from the $V_\tau$ component of
the pentagon condition.  So there is no new information in the $V_1$
component.

We can, however, get some additional information if we impose the
requirement that the associativity isomorphisms be unitary
transformations.  This amounts to requiring the vector spaces of
morphisms $\Hom(X,Y)$ to be Hilbert spaces and requiring our natural
bases for them to be orthonormal.

Unitarity tells us nothing new about $p$, since we already know $p=1$,
but unitarity of $
\begin{pmatrix}
  q&r\\s&t
\end{pmatrix}$ gives the equations
\[
q^2+|r|^2=q^2+|s|^2=1 \quad\text{and}\quad
q(\bar s-r)=q(s-\bar r)=0,
\]
where bars denote complex conjugation and where we used the fact that
$q$ is real.  So $s=\bar r$ and, since $rs=q$, we get first that
$q$ has to be positive,
\[
q=\frac{-1+\sqrt5}2,
\]
and second that
\[
r=\sqrt q e^{i\theta} \quad\text{and}\quad s=\sqrt q e^{-i\theta}
\]
for some real $\theta$.  Thus, we finally have, under the assumption
of unitarity,
\[
\alpha_{\tau,\tau,\tau}=
\begin{pmatrix}
  1&0&0\\0&q&\sqrt q e^{i\theta} \\0&\sqrt q e^{-i\theta}&-q
\end{pmatrix}
\]
with $q=\frac{-1+\sqrt5}2$ and $\theta$ arbitrary.  The presence of
$\theta$ here is an artifact of our choice of bases.  If we modified
the final vector in each of our bases, $\two {\two \tau\tau}\tau$ in
the domain of $\alpha_{\tau,\tau,\tau}$ and $\two \tau{\two \tau\tau}$
in the codomain, by a phase factor $e^{-i\theta}$, then, with respect
to the new bases, we would have
\[
\alpha_{\tau,\tau,\tau}=
\begin{pmatrix}
  1&0&0\\0&q&\sqrt q  \\0&\sqrt q &-q
\end{pmatrix}.
\]

\subsection{Braiding}
We now turn to the task of computing the braiding $\sigma$ in the
Fibonacci anyon category \scr A.  The only nontrivial component of the
natural isomorphism $\sigma$ is $\sigma_{\tau,\tau}$, because
components with a subscript 1 are identity morphisms and components
with non-simple subscripts reduce to direct sums of components with
simple subscripts.

The nontrivial component $\sigma_{\tau,\tau}$ is an isomorphism from
$\tau\otimes\tau=1\oplus\tau$ to itself.  Representing objects of \scr
A by pairs of vector spaces, we have that $\sigma_{\tau,\tau}$ is an
automorphism of $(\bbb C,\bbb C)$, so it amounts to two non-zero
scalars, $a$ multiplying vectors in the first (1) component and $b$
multiplying vectors in the second ($\tau$) component.  These are
subject to the hexagon identity, which equates the composites

\medskip
\[
\xymatrix{
& \tau\otimes(\tau\otimes \tau)
  \ar[0,1]^{\textstyle{\sigma}_{\tau,\tau\otimes \tau}}
& (\tau\otimes \tau)\otimes \tau
  \ar[1,1]^{\textstyle{\alpha}_{\tau,\tau,\tau}}
\\
(\tau\otimes \tau)\otimes \tau
  \ar[-1,1]^{\textstyle{\alpha}_{\tau,\tau,\tau}}
  \ar[1,1]_{\textstyle{\sigma}_{\tau,\tau}\otimes I_\tau}
&&& \tau\otimes (\tau\otimes \tau)
\\
& (\tau\otimes \tau)\otimes \tau
  \ar[0,1]_{\textstyle{\alpha}_{\tau,\tau,\tau}}
& \tau\otimes(\tau\otimes \tau)
  \ar[-1,1]_{\textstyle{I_\tau}\otimes \textstyle{\sigma}_{\tau,\tau}}
}
\]
\medskip

\noindent as well as the analogous identity with $\sigma^{-1}$ in place of
$\sigma$.

Consider the first (1) component of this equation.  In the bottom
composition, the $\sigma_{\tau,\tau}$ factors in the first and third
morphisms must act on the $\tau$ components so that the
$\otimes$-product with $I_\tau$ has a 1 component.  So both of these
are $b$.  The $\alpha$ between them, acting on the 1 component, is an
identity map, because our previous calculation gave $p=1$.  So the
bottom of the hexagon is $b^2$.  In the top, both of the $\alpha$'s
are again just 1.  The $\sigma$ in the middle of that row is
$\sigma_{\tau,1\oplus\tau}$, i.e., the direct sum of $\sigma_{\tau,1}$
and $\sigma_{\tau,\tau}$.  The first of these two summands has no 1
component; the second does, and it is $a$.  So the top of the hexagon
is just $a$, and the hexagon condition reads $a=b^2$.  (The
corresponding calculation for $\sigma^{-1}$ gives only
$a^{-1}=b^{-2}$, which is no new information.)

Now consider the second ($\tau$) component of the hexagon equation.
We do the calculation in matrix form, using the natural bases
\[
\two {\one \tau\tau}\tau \quad\text{and}\quad \two {\two \tau\tau}\tau
\quad\text{for }(\tau\otimes\tau)\otimes\tau
\]
and
\[
\two \tau{\one \tau\tau} \quad\text{and}\quad \two \tau{\one \tau\tau}
\quad\text{for }\tau\otimes(\tau\otimes\tau).
\]
With respect to these bases, $\alpha_{\tau,\tau,\tau}$ is given by $
\begin{pmatrix}
  q&r\\s&t
\end{pmatrix}$ as computed earlier.  Both $\sigma_{\tau,\tau}\otimes
I_\tau$ and $I_\tau\otimes\sigma_{\tau,\tau}$ are given by
\[
\begin{pmatrix}
  a&0\\0&b
\end{pmatrix}=
\begin{pmatrix}
  b^2&0\\0&b
\end{pmatrix},
\]
because in each case, $\sigma_{\tau,\tau}$ acts as $a$ on the first
basis vector (where it interchanges two $\tau$'s that were combined to
1) and as $b$ on the second (where it interchanges two $\tau$'s that
were combined to $\tau$).  Finally, $\sigma_{\tau,\tau\otimes\tau}$ is
the direct sum of $\sigma_{\tau,1}$ which is 1 and
$\sigma_{\tau,\tau}$ acting on the $\tau$ component, which is $b$;
since that direct sum decomposition matches our choice of bases,
$\sigma_{\tau,\tau\otimes\tau}$ is given by the matrix $
\begin{pmatrix}
  1&0\\0&b
\end{pmatrix}$.  Multiplying the matrices for each of the rows, we
find that the hexagon identity, in the $\tau$ component, reads
\[
\begin{pmatrix}
  q^2+brs&(q+bt)r\\
(q+bt)s&rs+bt^2
\end{pmatrix}=
\begin{pmatrix}
  b^4q&b^3r\\
b^3s&b^2t
\end{pmatrix}.
\]
Since we know, from our associativity calculation, that $r$ and $s$
are not zero, the (1,2) and (2,1) entries of this matrix equation
reduce to $q+bt-b^3=0$, or, since $t=-q$,
\[
b^3=q(1-b).
\]
The (1,1) and (2,2) entries give, after we remember that $rs=q$ and
cancel a common factor $q$,
\[
q+b=b^4\quad\text{and}\quad 1+bq+b^2=0.
\]
The last of these equations, being quadratic in $b$, can be solved
explicitly:
\[
b=\frac{-q\pm\sqrt{q^2-4}}2.
\]
We note that, since $q=\frac{\sqrt5-1}2$ is between 0 and 1, the
square root in the formula for $b$ is imaginary, so the two values of
$b$ are each other's complex conjugates.  The product of the two
values for $b$ is 1, so $b$ is a complex number of absolute value 1
with real part $\frac{-q}2$.

The ambiguity in the choice of $b$ is unavoidable in this situation.
Replacing one choice by the other just replaces $\sigma$ by its
inverse (since $|b|=1$), and there is nothing in the algebra of \scr A
that distinguishes the counterclockwise motion defining $\sigma$ from
the clockwise motion defining $\sigma^{-1}$.  To put it another way,
the change from one value of $b$ to the other can be exactly
compensated by reflecting the orientation of the (2-dimensional) space
in which the anyons live.

Although we have now computed $b$ and thus also $a=b^2$, we can get a
more useful view of these numbers by manipulating the three equations
above that relate $b$ to $q$.  Solving the last one for $q$ in terms
of $b$, and substituting the result, $q=\frac{-b^2-1}b$ into the other
two equations, we obtain from the first  equation that
\[
b^3=\frac{b^3-b^2+b-1}b,\text{ i.e., }
b^4-b^3+b^2-b+1=0,
\]
which means that $-b$ is a primitive fifth root of unity and therefore
$b$ is a primitive tenth root of unity.  The third equation above
confirms that by reducing to $b^5=-1$.

Among the four primitive tenth roots of unity only two, $e^{\pm3\pi
  i/5}$, have negative real parts, as $b$ does (recall that its real
part is $-q/2$).  So we conclude that, up to complex conjugations,
\[
b=e^{3\pi i/5} \text{ and therefore }a=e^{6\pi i/5}.
\]
This completes the calculation of the braiding $\sigma$ for Fibonacci
anyons.

\begin{rmk}
  The multiplicative structure for Fibonacci anyons, summarized by the
  fusion rule $\tau\otimes\tau=1\oplus\tau$, is perhaps the simplest
  nontrivial fusion rule.  Other fusion rules have been analyzed,
  either by hand as we have done here or with computer support.  The
  appendix of \cite{bonder} summarizes much of what is known about
  specific examples.  There does not, however, seem to be any general
  theory for arbitrary fusion rules.
\end{rmk}

\subsection{Fibonnaci Anyons and Quantum Computation}

In Section~\ref{anyons}, we mentioned the hope that, by using anyons
to encode qubits, one could use braiding to transform anyon states in
various ways, thereby enabling quantum computation.  Two anyons are
not sufficient for this purpose, because the braid group on two
strands is abelian, whereas quantum computation needs non-commuting
unitary transformations.  In the case of Fibonacci anyons, the
computation in the preceding subsection shows that the braiding
transformation $\sigma_{\tau,\tau}$ is diagonal in a suitable basis,
so it splits into one-dimensional representations; this again shows
its inadequacy for quantum computation.

With three Fibonacci anyons, the situation improves dramatically.  In
a suitable basis, the transformation that braids the first two of the
three anyons, $\sigma_{\tau,\tau}\otimes I_\tau$, is still diagonal.
The same goes for  the transformation that braids the second and third
anyons, but the suitable  bases in these two cases are not the same.
They differ by an associativity isomorphism $\alpha$.  More precisely,
one is the conjugate of the other by $\alpha_{\tau,\tau,\tau}$.  They
do not commute.

In fact, such braiding transformations suffice to approximate arbitrary
unitary transformations of the two-dimensional Hilbert space $V_\tau$
for $\tau^{\otimes 3}$.  Furthermore, using six Fibonacci anyons to
code two qubits, one can approximate, by braiding, the so-called
``controlled not'' gate, which, in combination with one-qubit gates,
is sufficient to produce all unitary gates for an arbitrary number of
qubits; that is, it is sufficient for quantum computation.  We refer
to \cite[Section~6]{panang} for these combinations of Fibonacci
braidings.


\begin{thebibliography}{99}

\bibitem{b-m}
Bhaskar Bagchi and Gadadhar Misra, ``A note on the multipliers and
projective representations of semi-simple Lie groups,''
\emph{Sankhy\=a: The Indian Journal of Statistics} 62 (2000)425--432.

\bibitem{bonder}
Parsa Hassan Bonderson, \emph{Non-Abelian Anyons and Interferometry,}
Ph.D. thesis, California Institute of Technology (2007).

\bibitem{freyd}
Peter Freyd, \emph{Abelian Categories: An Introduction to the Theory
  of Functors,} Harper \&\ Row (1964).

\bibitem{giulini} Domenico Giulini, ``Superselection rules''
  \url{http://arxiv.org/pdf/0710.1516.pdf}

\bibitem{kl}
  Louis Kauffman and Samuel Lomonaco, ``Topological Quantum
  Information Theory,'' in \emph{Quantum Information Science and Its
    Contribution to Mathematics,} ed. by S. Lomonaco,
  Proc. Symp. Appl. Math. 68, American Mathematical Society (2010)
  103--177.

\bibitem{kitaev}
Alexei Kitaev, ``Fault-tolerant quantum computation by anyons,''
\emph{Ann. Phys.} 303 (2003) 2--30.
\url{http://arxiv.org/abs/quant-ph/9707021}

\bibitem{maclane}
Saunders Mac Lane, \emph{Categories for the Working Mathematician},
Springer-Verlag, Graduate Texts in Mathematics 5 (1971).

\bibitem{nssfd}
Chetan Nayak, Steven Simon, Ady Stern, Michael Freedman, and Sankar
Das Sarma, ``Non-abelian anyons and topological quantum computation,''
\emph{Rev. Modern Phys.} 80 (2008) 1083-1159.

\bibitem{panang}
Prakash Panangaden and \'Eric O. Paquette, ``A categorical presentation of
quantum computation with anyons,'' Chapter~15  in \emph{New Structures
for Physics,} ed. by B. Coecke, Springer-Verlag Lecture Notes in
Physics 813 (2011) 983--1025.

\bibitem{Raghu}
Madabusi S. Raghunathan, ``Universal central extensions,''
\emph{Rev. Math. Phys.} 6 (1994) 207--225.

\bibitem{wang}
Zhenghan Wang, \emph{Topological Quantum Computation,} CBMS Regional
Conference Series in Mathematics, vol.~112, American Mathematical
Society (2010).


\end{thebibliography}
\end{document}